\documentclass[10pt]{article}
\usepackage{graphicx} 
\usepackage{amsmath}
\usepackage{amssymb}
\usepackage{cite}

\title{Little Rip, Pseudo Rip and bounce cosmology with generalized equation of state in non-standard backgrounds}
\author{Oem Trivedi$^a$ \footnote{oem.t@ahduni.edu.in}  \and Alexander V. Timoshkin $^{b,c}$ \footnote{alex.timosh@rambler.ru} }
\date{%
	$^a$School of Arts and Sciences, Ahmedabad University, Ahmedabad 380009,India\\%
	$^b$Institute of Scientific Research and Development,Tomsk State Pedagogical University (TSPU), 634061 Tomsk, Russia\\%
       $^c$ Lab. for Theor. Cosmology, International Centre of Gravity and Cosmos,
Tomsk State University of Control Systems and Radio Electronics (TUSUR), 634050 Tomsk, Russia\\%
	\today
}

\begin{document}

\maketitle

\begin{abstract}
    The discovery of universe's late-time acceleration and dark energy has overseen a great deal of research into cosmological singularities and particularly future singularities. Perhaps the most extreme of such singlarities is the big rip, which has propelled a lot of work into ways of moderating it or seeking out alternatives to it and two such alternatives to the big rip are the Little rip and Pseudo rip. Another possibility to consider the far future of the universe is through bounce cosmologies, which presents its own interesting ideas. So in this work we investigate the Little rip, Pseudo rip and Bounce cosmology in non-standard cosmological backgrounds with a generalized equation of state in the presence of a viscous fluid. In particular we discuss about Chern-Simons cosmology and the RS-II Braneworld and discuss how the exotic and non-conventional nature of gravity in such cosmologies affect universal evolution in these scenarios. We find out that there are very significant differences in the behaviour of such cosmic scenarios in these universes in comparison to how they appear in the simple general relativistic universe.  
\end{abstract}

\section{Introduction}
The surprising observation of the universe's accelerated expansion during its later stages has prompted extensive research within the field of cosmology \cite{SupernovaSearchTeam:1998fmf}.Numerous approaches have been explored to explain this expansion, ranging from conventional methods like the Cosmological constant \cite{SupernovaSearchTeam:1998fmf,Weinberg:1988cp,Lombriser:2019jia,Copeland:2006wr,Padmanabhan:2002ji} to more exotic scenarios such as Modified gravity theories \cite{Capozziello:2011et,Nojiri:2010wj,Nojiri:2017ncd} and recent proposals for direct detection of dark energy \cite{Zhang:2021ygh}. It is widely recognized that a dark energy universe exhibits peculiar characteristics regarding singularities in the distant future, even in scenarios where it arises from modified gravity \cite{Nojiri:2006ri,Nojiri:2003ft,Nojiri:2004ip,Nojiri:2004pf,Nojiri:2005sr,Nojiri:2006zh}. Among the various models investigated in the literature, there are those that treat dark energy and dark matter as ideal (nonviscous) fluids with an unconventional equation of state (EoS). General dark fluid models with an inhomogeneous EoS were introduced in \cite{Brevik:2011mm,Brevik:2019mah,Brevik:2022aqp}
.
\\
\\
Numerous cosmological scenarios exist for the evolution of the universe, including the Big Rip \cite{Caldwell:2003vq,Nojiri:2004ip}, the Little Rip \cite{Frampton:2011sp,Brevik:2011mm,Frampton:2011rh,Astashenok:2012kb,Astashenok:2012ps,Nojiri:2011kd,Makarenko:2012gm} , the Pseudo Rip \cite{Frampton:2011aa}, and the Quasi Rip \cite{Wei:2012ct}. The analytical representation of the cosmological models Little Rip and Pseudo Rip , in terms of the parameters of the generalized equation of state without considering the bulk viscosity of the dark fluid in flat space, was derived in previous articles. The investigation of bounce cosmological models in the flat FRW metric, considering the viscosity properties of a dark fluid, was conducted in another work and reconstructions within the framework of modified gravity theories have also been done \cite{Brevik:2012qy,Brevik:2011mm,Brevik:2012ka,Brevik:2013sea,Brevik:2013qka,Brevik:2019mah,Bamba:2008ut,Bamba:2010wfw,Bamba:2012vg}.
Therefore, the equation of state plays a significant role not only in the Little Rip, Pseudo Rip, and bounce phenomena but also in other cosmological scenarios. Analytical expressions for the spatial curvature, thermodynamic parameter, and bulk viscosity in these models are obtained. Specifically, the parameters for the early and late stages of the universe's history are derived.
\\
\\
While astronomical measurements of the luminosity of remote objects using the Planck satellite indicate that our universe is nearly flat \cite{Planck:2018nkj}, other studies such as observations of galaxy distribution in space challenge this conclusion. It could be a viable possibility that the universe could eventually possess finite spatial curvature\cite{DiValentino:2019qzk,Handley:2019tkm} and so investigations that consider the combined influence of bulk viscosity and spatial curvature to describe the evolution of the universe have gathered interest. A lot of work has been dedicated to studying singularities and their ultimate formation time. However, singularities are not the only possible endpoints of the universe's evolution. The cosmological models of the late-time universe, including the Little Rip, Pseudo Rip, and the bounce cosmology describing a cyclic universe, as discussed in this article, are nonsingular. Hence recent work explored the Little Rip, Pseudo Rip, and the bounce cosmological scenarios in cosmologies with viscous fluids and discussed their properties in detail \cite{Timoshkin:2023pfb}. Here we would like to go a step further and discuss these scenarios in non-conventional cosmological paradigms, namely the RS-II Braneworld \cite{Randall:1999ee,Randall:1999vf} and a form of cosmology in Chern-Simons gravity \cite{gomez2011standard}. We choose these paradigms as they are both inspired heavily by quantum gravitational ideas and also represent first-order corrections to cosmology due to extra-dimensional considerations. So it makes for an interesting endeavor to probe such scenarios in these cosmologies and note any differences that they bring out in these paradigms from the usual general relativistic cosmology. In Section 2, we would briefly review the viscous fluid regime we would be probing, and in Section 3, we would discuss both forms of rip and the bounce scenario in these cosmologies. We would then conclude the work in Section 4.
\\
\\
\section{ Overview of the cosmologies considered}
Here we shall briefly discuss the viscous fluid cosmological scenario that we are interested in this work. We are examining a universe that is filled with a one-component ideal fluid known as dark energy, in the framework of the FRW metric. This universe possesses a nonzero spatial curvature and is described by the scale factor $a$.
\\
\\
Let's express the energy conservation law as follows:
\begin{equation}
\dot{\rho}+3H(\rho+p)=0. \label{1}
\end{equation}
Here, $H=\dot{a}/a$ represents the Hubble rate, and $k^2=8\pi G$ where $G$ denotes Newton's gravitational constant. The quantities $\rho$ and $p$ correspond to the energy density and pressure of the dark energy, respectively. The dot notation signifies differentiation with respect to cosmic time $t$.
\\
\\
We will now investigate the homogeneous and isotropic FRW expanding universe, given by the metric:
\begin{equation}
ds^2 = -dt^2 + a^2(t)\left( \frac{dr^2}{1-\Pi_0 r^2}+r^2 d\Omega^2 \right), \label{2}
\end{equation}
In this equation, $d\Omega^2= d\theta^2+\sin^2\theta d\varphi^2$, $t$ represents cosmic time, $a(t)$ represents the scale factor (with length units), $r$ is the special radius coordinate, and $\Pi_0$ signifies the spatial curvature of the three-dimensional space.
\\
\\
As we know, Eq.~(\ref{1}) geometrically describes various types of universes. Assuming $\Pi_0$ to be dimensionless for simplicity, when $\Pi_0 = 0$, the universe is spatially flat; for $\Pi_0 = 1$, it is closed; and for $\Pi_0 = -1$, it is open. The nature of the universe's expansion depends on the spatial curvature: an open universe will expand indefinitely, a flat universe will also expand indefinitely but with a constant expansion velocity as $t \rightarrow +\infty$, and a closed universe will expand until a certain moment, after which the expansion will transition to compression, eventually leading to a collapse.
\\
\\
To describe an inhomogeneous viscous fluid, we adopt the following equation of state (EoS) \cite{Nojiri:2006zh}:
\begin{equation} \label{contequation}
p= \omega(\rho,t)\rho-3H\zeta(H,t)
\end{equation}
Here, $\zeta(H,t)$ represents the bulk viscosity, which depends on $H$ and $t$. From conventional thermodynamics, we know that $\zeta(H,t)>0$. We assume the EoS parameter $\omega$ to have the form \cite{Nojiri:2006zh}:
\begin{equation}
\omega(\rho,t)=\omega_1(t)(A_0\rho^{\alpha-1} -1), \label{5}
\end{equation}
In this equation, $A_0 \neq 0$ and $\alpha \geq 1$ is a constant. Regarding dimensions, since $\omega$ and $\omega_1$ are dimensionless, the dimension of $A_0$ becomes complicated when $\alpha >1$. However, in the simplest case of $\alpha =1$, $A_0$ becomes dimensionless. Consequently, we set $\omega(\rho,t)=\omega_0$ as a constant.Dissipative processes are described by the bulk viscosity in the form \cite{Nojiri:2006zh}:
\begin{equation}
\zeta(H,t)= \zeta_1(t)(3H)^n, \label{6}
\end{equation}
Here, $n>0$ and $\zeta_1(t)$ is an arbitrary function dependent on time.We will now explore cosmological models of the viscous fluid.
\\
\\
Furthermore the Little rip scenario we will consider is specified by the following form for the Hubble parameter \cite{Frampton:2011rh,Frampton:2011sp}:
\begin{equation}
H=H_0 \exp(\lambda t), \quad H_0>0, \lambda >0. \label{7}
\end{equation}
We will follow the development of the universe beginning at some time $t=0$ which will at first be left unspecified,
except that it refers to an initial instant in the very early universe. Its precise meaning will be dependent on which model we consider.
The symbol  $H_0$ means the Hubble parameter at this particular instant. The Pseudo Rip, proposed in \cite{Frampton:2011aa}, is a second variant of the Little Rip.
This interesting possible scenario is related to the LR cosmology when the
Hubble parameter tends to infinity in the remote future
\begin{equation}
H(t)\rightarrow H_\infty, \quad t\rightarrow +\infty. \label{14}
\end{equation} The pseudo rip scenario will also be specified by an ansatz for the Hubble parameter, which will be given in this case \cite{Frampton:2011aa} \begin{equation}
H=H_0-H_1\exp(-\lambda t), \label{15}
\end{equation}
where $H_0, H_1$ and $\lambda$ are positive constants, $H_0>H_1$, and $t>0$. For more information on the Little and Pseudo rips, the reader can refer to \cite{Brevik:2011mm,Brevik:2021sgw,Brevik:2020nuo,Brevik:2015pda,Brevik:2014eya,Brevik:2012ka,lohakare2022rip,BorislavovVasilev:2021srn,BorislavovVasilev:2021elf,Astashenok:2012kb,Granda:2011kx,Brevik:2012nt}. 
In the case of bounce cosmology, the universe goes from an era of accelerated collapse to the expanding era without displaying
a singularity. This is essentially a cyclic universe model. After the bounce the universe soon enters a matter dominated
expansion phase \cite{Novello:2008ra,Bamba:2013fha,Cai:2011tc}. For this we can consider two ansatz, the first one where the scale factor $a(t)$ has an exponential form \cite{Bamba:2013fha}:

\begin{equation}
a(t)=a_0 \exp[\alpha(t-t_0)^{2n}], \quad n \in N, \label{expansatz}
\end{equation}
where $\alpha$ is a positive (dimensional) constant and $t_0$ is the instant when bouncing occurs.In this case the Hubble parameter becomes \begin{equation}
H(t)=2n\alpha(t-t_0)^{2n-1}. \label{21}
\end{equation}
Another viable ansatz for a bounce scenario is where the scale factor has the following form \cite{Bamba:2013fha}:
\begin{equation}
a(t)=a_0+\alpha(t-t_0)^{2n}, \label{powerlawansatz}
\end{equation}
where $\alpha$ is dimensional constant, $n \in N$ and $t=t_0$ is a fixed bounce time.
The Hubble parameter in this case becomes
\begin{equation}
H(t)=\frac{2n\alpha(t-t_0)^{2n-1}}{a_0+\alpha(t-t_0)^{2n}}. \label{27}
\end{equation}
We will now investigate all of these scenarios in a universe with a viscous fluid form of dark energy, in both the Chern-Simons cosmology and the RS-II Braneworld cosmology.  
\section{Chern Simons cosmology}
The Friedmann equation which we will be considering here is given by  that we will be covering is given by \begin{equation} \label{a1}
	\frac{6}{k} \left(\frac{\Pi_{0}}{a^2}+H^2\right)+\alpha \left(\frac{\Pi_{0}}{a^4}+\frac{2 H^2 \Pi_{0}}{a^2}+H^4\right) = \rho
	 \end{equation}
	 where in the context of the Chern-Simons scenario, k is a constant which is positive while $\alpha $ is also a parameter which is considered to be a constant but is not necessarily positive but we would consider it to be positive for our analysis for simplicity. The above equation is very special in the sense that it can be derived from various distinct approaches. This equation can be achieved by considering a quantum corrected entropy-area relation of $ S = \frac{A}{4} - \alpha \ln \left( \frac{A}{4} \right) $ where A is the area of the apparent horizon and $\alpha$ is a dimensionless positive constant determined by the conformal anomaly of the fields, where the conformal anomaly is interpreted as a correction to the entropy of the apparent horizon \cite{cai2008corrected} . Also, this could be derived in terms of spacetime thermodynamics together with a generalized uncertainly principle of quantum gravity\cite{lidsey2013holographic}. This Friedmann equation can also be derived by considering an anti-de Sitter-Schwarzschild black hole via holographic renormalization with appropriate boundary conditions \cite{apostolopoulos2009cosmology}. Finally, a Chern-Simons type of theory can also yield this Friedmann equation \cite{gomez2011standard}. Hence the equation (\ref{a1}) can derived by a wide range of approaches to gravitational physics and can be representative of the effects of these different theories on the cosmological dynamics.
     \\
     \\
     Our analysis will be based in the simplest case, when $\omega(\rho,t)=\omega_0$ is a constant, and $\zeta(H,t)=\zeta_0$, with a constant $\zeta_0>0$, for which the generalized equation of state takes the form: \begin{equation}
            p=\omega_0 \rho-3\zeta_0 H. \label{8}
            \end{equation}
    Considering the Friedmann equation (\ref{a1}), we can write the derivative of the energy density as \begin{equation}
    \dot{\rho} = \frac{12 H^2 \lambda \left(\alpha \left(\frac{\Pi_{0}}{a^2}+2 H^2\right)+1\right)}{k}
\end{equation}
As we would firstly like to discuss the Little rip scenario, using the ansatz (\ref{7}), we have the continuity equation as \begin{equation}
3 H \left(\frac{4 H \lambda \left(\frac{\alpha \Pi_{0}}{a^2}+2 \alpha h^2+1\right)}{k}+\frac{(w+1) \left(a^2 H^2+\Pi_{0}\right) \left(a^2 \left(\alpha H^2 k+6\right)+\alpha k \Pi_{0}\right)}{a^4 k}-3 \zeta_{0} H\right) = 0    
\end{equation}         
In the limit $t \to 0$, we have \begin{equation}
    w = -\frac{8 \alpha g H_{0}^2+\alpha H_{0}^3 k-3 \zeta_{0} k+4 \lambda+6 H_{0}}{\alpha H_{0}^3 k+6 H_{0}}
\end{equation}
In the limit $t \to \infty$, we have $w \to -1$ for all positive $\lambda$ . So in the very late universe in this case, we see pure de Sitter behaviour irrespective of anything. In the case $\Pi_{0} \neq 0 $, we have the w parameter \begin{equation}
    w = \frac{-8 a^4 \alpha \lambda H^3-a^4 \alpha H^4 k+3 a^4 \zeta_{0} H k-4 a^4 \lambda H-6 a^4 H^2-4 a^2 \alpha \lambda H \Pi_{0}-2 a^2 \alpha H^2 k \Pi_{0}-6 a^2 \Pi_{0}-\alpha k \Pi_{0}^2}{\left(a^2 H^2+\Pi_{0}\right) \left(a^2 \alpha H^2 k+6 a^2+\alpha k \Pi_{0}\right)}
\end{equation}
Using the ansatz, we have \begin{equation} \label{chernw}
    w = -1 -\frac{a^2 H_{0} e^{\lambda t} \left(a^2 \left(\lambda \left(8 \alpha H_{0}^2 e^{2 \lambda t}+4\right)-3 \zeta_{0} k\right)+4 \alpha \lambda \Pi_{0}\right)}{\left(a^2 H_{0}^2 e^{2 \lambda t}+\Pi_{0}\right) \left(a^2 \left(\alpha H_{0}^2 k e^{2 \lambda t}+6\right)+\alpha k \Pi_{0} \right)}
\end{equation}
In the far future limit $t \to \infty$ we have $w\to -1$, pure de Sitter behaviour. In the limit $t \to 0$, we have \begin{equation}
    w = -1 -\frac{a^2 H_{0} \left(a^2 \left(\lambda \left(8 \alpha H_{0}^2+4\right)-3 \zeta_{0} k\right)+4 \alpha \lambda \Pi_{0}\right)}{\left(a^2 H_{0}^2+\Pi_{0} \right) \left(a^2 \left(\alpha H_{0}^2 k+6\right)+\alpha k \Pi_{0} \right)}
\end{equation}
In either case of non flat or flat universe, we have the universe tending towards neither a Quintessence nor a phantom behaviour in the far future limit, something which is not observed in the case of the conventional cosmology \cite{Timoshkin:2023pfb}.
\\
\\
The continuity equation in the case for the pseudo rip with $H(t) = H_{0} - H_{1} \exp(-\lambda t)$ will be given by
\begin{multline}
\frac{4 \lambda \mathrm{h}_1 e^{-4 \lambda t} \left(H_0 e^{\lambda t}-H_1\right) \left(a^2 \left(e^{2 \lambda t} \left(\alpha H_0^2 k+3\right)-2 \alpha H_0 H_1 k e^{\lambda t}+\alpha H_1^2 k\right)+\alpha k \Pi_0 e^{2 \lambda t}\right)}{a^2 k}+ \\ 3 \left(H_0-H_1 e^{-\lambda t}\right) \left((w+1) \left(\frac{6 \left(\frac{\Pi_0}{a^2}+\left(H_0-H_1 e^{-\lambda t}\right)^2\right)}{k}+\frac{\alpha e^{-4 \lambda t} \left(a^2 \left(H_1-H_0 e^{\lambda t}\right)^2+\Pi_0 e^{2 \lambda t}\right)^2}{a^4}\right) \right) - \\ 3 \left(H_0-H_1 e^{-\lambda t}\right) (3 \epsilon_0 \left(H_0-H_1 e^{-\lambda t}\right)) = 0
\end{multline}
This gives us the equation of state parameter as,
\begin{equation}
\begin{split}
w = -\frac{1}{3 \left(a^2 \left(H_{1}-H_{0} e^{\lambda t}\right)^2+\Pi_{0} e^{2 \lambda t}\right) \left(a^2 \left(e^{2 \lambda t} \left(\alpha H_{0}^2 k+6\right)-2 \alpha H_{0} H_{1} k e^{\lambda t}+\alpha H_{1}^2 k\right)+\alpha k \Pi_{0} e^{2 \lambda t}\right)} \\
\times \Bigg[ 4 a^4 \alpha H_{1}^3 k (\lambda-3 H_{0}) e^{\lambda t}+3 a^4 \alpha H_{1}^4 k+a^2 H_{1} e^{3 \lambda t} \left(a^2 \left(4 \alpha H_{0}^2 k (\lambda-3 H_{0})+9 (\zeta_{0} k-4 H_{0})+12 \lambda\right)+ \right. \\
\left. 4 \alpha k \Pi_{0} (\lambda-3 H_{0})\right)-2 a^2 H_{1}^2 e^{2 \lambda t} \left(a^2 (\alpha H_{0} k (4 \lambda-9 H_{0})-9)-3 \alpha k \Pi_{0}\right)+3 e^{4 \lambda t} \left(a^4 H_{0} \left(\alpha H_{0}^3 k-3 \zeta_{0} k+6 H_{0}\right)+ \right. \\
\left. 2 a^2 \Pi_{0} \left(\alpha H_{0}^2 k+3\right)+\alpha k \Pi_{0}^2\right) \Bigg]
\end{split}
\end{equation}
For $\Pi_{0} = 0$, we have in the $t \to 0$ limit
\begin{equation}
w = \frac{(H_{0}-H_{1}) \left(-\alpha k (H_{0}-H_{1}) \left(4 \lambda H_{1}+3 (H_{0}-H_{1})^2\right)+9 \zeta_{0} k+18 (H_{1}-H_{0})\right)-12 \lambda H_{1}}{3 (H_{0}-H_{1})^2 \left(\alpha k (H_{0}-H_{1})^2+6\right)}
\end{equation}
While in the far-future limit, we have
\begin{equation} \label{eq13}
w = -1 -\frac{3 \zeta_{0} k}{\alpha H_{0}^3 k+6 H_{0}}
\end{equation}
Here we see that similar to the case of the conventional cosmology, in the far future the phantom regime will take over in this case even for a Chern Simons cosmology. Finally, the bulk viscosivity parameter in this case is given as \begin{multline}
    \zeta_{0} = \frac{\frac{4 H_{1} \lambda  e^{-4 \lambda  t} \left(H_{0} e^{\lambda  t}-H_{1}\right) \left(a^2 \left(e^{2 \lambda  t} \left(\alpha  H_{0}^2 k+3\right)-2 \alpha  H_{0} H_{1} k e^{\lambda  t}+\alpha  H_{1}^2 k\right)+\alpha  k \text{$\pi $0} e^{2 \lambda  t}\right)}{a^2 k}}{9 \left(H_{0}-H_{1} e^{\lambda  (-t)}\right)^2} + \\ \frac{3 (w+1) \left(H_{0}-H_{1} e^{\lambda  (-t)}\right) \left(\frac{6 \left(\frac{\text{$\pi $0}}{a^2}+\left(H_{0}-H_{1} e^{\lambda  (-t)}\right)^2\right)}{k}+\frac{\alpha  e^{-4 \lambda  t} \left(a^2 \left(H_{1}-H_{0} e^{\lambda  t}\right)^2+\text{$\pi $0} e^{2 \lambda  t}\right)^2}{a^4}\right)}{9 \left(H_{0}-H_{1} e^{\lambda  (-t)}\right)^2} 
\end{multline} 
We see that the viscosity parameter is influenced by the value of the curvature $\Pi_{0} $ very much, just like it is in the case of the conventional cosmology as well.
\\
\\
Now we discuss the case of bounce cosmology, where we have for the exponential ansatz $a(t)=a_{0} \exp \left(\alpha (t-t_{0})^{2 n}\right)$ the continuity equation as
\begin{align}
&-4 \left(2 \alpha n (t-t_{0})^{2 n} \left(-2 a_{0} n+a_{0}+a_{1} (t-t_{0})^{2 n}\right)+p (t-t_{0})^2\right) \nonumber \\
&\quad \times \left(3 a_{0}^2 (t-t_{0})^2+6 a_{0} \alpha (t-t_{0})^{2 n+2} \right. \nonumber \\
&\quad \quad +\alpha \left(k \left(4 \alpha^2 n^2 (t-t_{0})^{4 n}+p (t-t_{0})^2\right)+3 \alpha (t-t_{0})^{4 n+2}\right)\bigg) \nonumber \\
&\quad +3 (w+1) \nonumber \\
&\quad \times \left(6 (t-t_{0})^4 \left(a_{0}+a_{1} (t-t_{0})^{2 n}\right)^2 \right. \nonumber \\
&\quad \quad \left. \times \left(4 a_{1}^2 n^2 (t-t_{0})^{4 n-2}+ \Pi_{0}  \right)+a_{1} k \left(4 a_{1}^2 n^2 (t-t_{0})^{4 n}+ \Pi_{0} (t-t_{0})^2\right)^2\right) \nonumber \\
&\quad -18 \alpha f k n (t-t_{0})^{2 n+3} \left(a_{0}+\alpha (t-t_{0})^{2 n}\right)^3 = 0
\end{align}

The expressions for the equation of state parameter and the viscosity parameter are awfully long in this case but we can still discuss some important issues here. Taking $n=1$, we have the EOS parameter at the time of bounce $t-> t_{0} $ as 
\begin{equation} \label{wb1}
    w = \frac{-24 a_{0}^3 a_{1}-6 a_{0}^2 \Pi_{0}-8 \alpha  a_{0} a_{1} k \Pi_{0}+\alpha  k \Pi_{0}^2}{18 a_{0}^2 \Pi_{0}+3 \alpha  k \Pi_{0}^2} 
\end{equation}
We see that for the case of the flat universe \eqref{wb1} blows up, which is similar to what happens in the same case for the conventional cosmology \cite{Timoshkin:2023pfb}. While in the case of the open universe, $\Pi = -1$, we have \begin{equation} \label{wb11}
    w = \frac{4 a_{0}^2 (2 a_{0} a_{1}+1)}{\alpha  k-6 a_{0}^2}+\frac{8 \text{a0} a_{1}}{3}+\frac{1}{3}
\end{equation}  
which shows that in this case $w > 0$ while for the case of the the closed universe \begin{equation}
   w= \frac{1}{3} -\frac{4 a_{0}^2}{6 a_{0}^2+\alpha  k}-\frac{24 a_{0}^3 a_{1}+8 \alpha  a_{0} a_{1} k}{18 a_{0}^2+3 \alpha  k}
\end{equation} In the usual form of cosmology, when one considers the case of non-flat universes one gets $w \to -1 $ but we see that in the case of the Chern Simons scenario that is not the case.
\\
\\
Considering now the power law bounce model with the ansatz \eqref{powerlawansatz}, we have the continuity equation as \begin{multline}
-4 \left(2 n a_{1} (t-t_{0})^{2 n} \left(-2 a_{0} n+a_{0}+a_{1} (t-t_{0})^{2 n}\right)+ \Pi_{0} (t-t_{0})^2\right) \left(3 a_{0}^2 (t-t_{0})^2+6 a_{0} a_{1} (t-t_{0})^{2 n+2}\right. \\
+\alpha k \left(4 n^2 a_{1}^2 (t-t_{0})^{4 n}+ \Pi_{0} (t-t_{0})^2\right)+3 a_{1}^2 (t-t_{0})^{4 n+2}\bigg)+3 (w+1) \\ \bigg(6 (t-t_{0})^4 \left(a_{0}+ a_{1} (t-t_{0})^{2 n}\right)^2 \left(4 n^2 a_{1}^2 (t-t_{0})^{4 n-2}\right)
+ \left. \Pi_{0}\right) +\alpha k \left(4 n^2 a_{1}^2 (t-t_{0})^{4 n} + \Pi_{0} (t-t_{0})^2\right)^2\bigg) - \\ 18 \zeta_{0} k n a_{1} (t-t_{0})^{2 n+3} \left(a_{0}+ a_{1} (t-t_{0})^{2 n}\right)^3 =0
\end{multline} 
From which we can get the equation of state parameter and the viscosity parameter as 
\begin{align}
\label{eq18}
w &= -1 + \frac{6\zeta_{0} k n a_{1}(t - t_{0})^{2n+3} \left(a_{0} + a_{1}(t - t_{0})^{2n}\right)^3}{6(t - t_{0})^4 \left(a_{0} + a_{1}(t - t_{0})^{2n}\right)^2 \left(4n^2a_{1}^2(t - t_{0})^{4n-2} + \Pi_{0}\right) + \alpha k \left(4n^2a_{1}^2(t - t_{0})^{4n} + \Pi_{0}(t - t_{0})^2\right)^2} \nonumber \\
&\quad + \frac{1}{3 \left(6(t - t_{0})^4 \left(a_{0} + a_{1}(t - t_{0})^{2n}\right)^2 \left(4n^2a_{1}^2(t - t_{0})^{4n-2} + \Pi_{0}\right) + \alpha k \left(4n^2a_{1}^2(t - t_{0})^{4n} + \Pi_{0}(t - t_{0})^2\right)^2\right)} \nonumber \\
&\times \Bigg[ 4 \left(2n a_{1}(t - t_{0})^{2n} \left(-2a_{0}n + \left( a_{0} + a_{1}(t - t_{0})^{2n} \right) \right) + \Pi_{0}(t - t_{0})^2\right) \nonumber \\
&\quad \times \left((t - t_{0})^2 \left(3a_{0}^2 + \alpha k \Pi_{0}\right) + 6a_{0}a_{1}(t - t_{0})^{2n+2} + a_{1}^2(t - t_{0})^{4n} \left(4\alpha kn^2 + 3(t - t_{0})^2\right)\right) \Bigg]
\end{align}
For the case of flat space $\Pi_{0}= 0$, we have \begin{multline} \label{wflatchernpower}
    w = -1 + \frac{3 a_{1}^3 (t-t_{0})^2 (3 \zeta_{0} k (t-t_{0})-8 n+4)+3 a_{0}^2 a_{1} (t-t_{0})^{2 n+2} (9 \zeta_{0} k (t-t_{0})-16 n+12)}{36 a_{0}^2 a_{1} n (t-t_{0})^{2 n+2}+72 a_{0} a_{1}^2 n (t-t_{0})^{4 n+2}+12 a_{1}^3 n (t-t_{0})^{6 n} \left(2 \alpha  k n^2+3 (t-t_{0})^2\right)} + \\ \frac{a_{0} a_{1}^2 (t-t_{0})^{4 n} \left(16 \alpha  k n^2 (1-2 n)+27 \zeta_{0} k (t-t_{0})^3-12 (2 n-3) (t-t_{0})^2\right)}{36 a_{0}^2 a_{1} n (t-t_{0})^{2 n+2}+72 a_{0} a_{1}^2 n (t-t_{0})^{4 n+2}+12 a_{1}^3 n (t-t_{0})^{6 n} \left(2 \alpha  k n^2+3 (t-t_{0})^2\right)} \\ \frac{a_{1}^3 (t-t_{0})^{6 n} \left(16 \alpha  k n^2+3 (t-t_{0})^2 (3 \zeta_{0} k (t-t_{0})+4)\right)}{36 a_{0}^2 a_{1} n (t-t_{0})^{2 n+2}+72 a_{0} a_{1}^2 n (t-t_{0})^{4 n+2}+12 a_{1}^3 n (t-t_{0})^{6 n} \left(2 \alpha  k n^2+3 (t-t_{0})^2\right)}
\end{multline}
The important thing to note is that the equation of state parameter for both \eqref{eq18} and \eqref{wflatchernpower} diverges at the time of the bounce $t \to t_{0}$. Such a blow up of the w parameter is also seen in the conventional cosmology, but only for the flat space case but interestingly for the Chern-Simons scenario this is also happening for the non-flat case too.  
\section{RS-II Braneworld}
The RS-II model is founded on a modification of the RS-I Braneworld cosmology model \cite{Randall:1999ee}. In the RS-I model, the hierarchy problem is addressed by embedding two 3-branes within a five-dimensional bulk, with one of these branes housing the Standard Model particles. In contrast, the RS-II braneworld cosmology simplifies the model by eliminating one of the 3-branes and successfully recovering both Newtonian gravity and General Relativity as its limiting cases \cite{Randall:1999vf}.

Given the potential impact of other braneworld cosmology scenarios, such as the Dvali-Gabadadze-Porrati (DGP) model \cite{Deffayet:2000uy, Dvali:2000hr}, on the late-time evolution of the universe, it becomes intriguing to explore whether the RS-II model can also address issues related to late-time acceleration. In recent times, several studies have delved into the topic of Quintessence within the RS-II Braneworld framework \cite{Sahni:2002dx, Sami:2004xk, Bento:2008yx}.

We can express the complete action of the RS-II Model, incorporating the background fluid terms too, as follows:
\begin{equation}
    S = S_{RS} + S_{B} = \int d^5 x \sqrt{-g^{(5)}} \left( \Lambda^{(5)} + 2 R^{(5)} \right) + \int d^4 x \sqrt{-g} \left( \omega + \mathcal{L}_{B} \right)
\end{equation}
Here, $R^{(5)}$, $g^{(5)}_{\mu \nu}$, and $\Lambda^{(5)}$ denote the bulk Ricci Scalar, the metric of the five-dimensional bulk, and the cosmological constant in that bulk, respectively. Additionally, $\omega$ represents the brane tension on the 3-brane, while $g_{\mu \nu}$ stands for the metric on the 3-brane.

If one wants to consider, for example, an RS-II scenario with scalar field dark energy then the action would be \begin{equation}
 S  = S_{RS} + S_{B} + S_{\phi} = \int d^5 x \sqrt{-g^{(5)}} \left( \Lambda^{(5)}  + 2 R^{(5)}   \right) + \int d^4 x \sqrt{-g} \left(\lambda -\frac{1}{2} \mu(\phi) (\nabla \phi)^2 - V(\phi)  + \mathcal{L}_{B}  \right) 
	\end{equation}
 with $\mu(\phi) $ being a scalar coupling function, but we are not restricting the cosmology to a scalar field dark energy scenario in any case. The Friedmann equation in this case for an FLRW background inclusive of the curvature term would be \begin{equation}
     H^2 = \frac{k^2}{3} \left( \rho + \frac{\rho }{3} \right) - \frac{\Pi_{0}}{a^2}
 \end{equation}
where k is again a positive term. dependent on the five dimensional Planck mass in this case $ m_{p}^{(5)} $. 
The energy density in this case can then be written as \begin{equation}
   \rho = \sqrt{\frac{6 \sigma \left(\frac{\Pi_{0}}{a^2}+H^2\right)}{k^2}+\sigma^2}-\sigma 
\end{equation}
And the continuity equation \eqref{contequation} can be written as \begin{equation}
    3 H \left(\sqrt{\sigma \left(\frac{6 \left(a^2 H^2+\Pi_{0}\right)}{a^2 k^2}+\sigma\right)} \left(\frac{2 a^2 H \lambda}{a^2 \left(\sigma k^2+6 H^2\right)+6 \Pi_{0}}+w+1\right)-b (w+1)-3 \zeta_{0} H\right) = 0
\end{equation}
In the case of flat space we can then write \begin{equation} \label{wrs2rip}
    w = -1 + \frac{\frac{3 \zeta_{0} k \left(\sqrt{\sigma \left(\sigma k^2+6 H^2\right)}+\sigma k\right)}{\sigma}+2 \lambda \left(-\frac{\sigma k}{\sqrt{\sigma \left(\sigma k^2+6 H^2\right)}}-1\right)}{6 H}
\end{equation}
This in the $t \to 0$ limit gives us \begin{equation}
    w = -1 + \frac{\frac{3 \zeta_{0} k \left(\sqrt{\sigma \left(\sigma k^2+6 H_{0}^2\right)}+\sigma k\right)}{\sigma}+2 \lambda \left(-\frac{\sigma k}{\sqrt{\sigma \left(\sigma k^2+6 H_{0}^2\right)}}-1\right)}{6 H_{0}}
\end{equation}
While in the far future limit $ t \to \infty $ we have \begin{equation} \label{wrs2farfuture}
    w = -1 + \frac{\sqrt{\frac{3}{2}} \zeta_{0} k}{\sqrt{\sigma}}
\end{equation}
Here we start to immediately see the stark difference provided by the RS II cosmology from what the Chern Simons and conventional GR cosmology provided us in the similar case. In the similar case for the conventional cosmology, one saw a phantom evolution $ w < -1 $ while for Chern Simons one saw a pure dS evolution but for the RS II braneworld, we see that here we will have a quintessence type evolution which is different from both the other scenarios. Moving further and talking about the non flat universe, we have for $\Pi_{0} \neq 0$ \begin{equation}
    w = \frac{-\frac{3 H \sqrt{\sigma \left(\frac{6 \left(a^2 H^2+\Pi_{0}\right)}{a^2 k^2}+\sigma\right)} \left(a^2 \left(\sigma k^2+2 H (3 H+\lambda)\right)+6 \Pi_{0}\right)}{a^2 \left(\sigma k^2+6 H^2\right)+6 \Pi_{0}}+3 \sigma H+9 \zeta_{0} H^2}{3 H \sqrt{\sigma \left(\frac{6 \left(a^2 H^2+ \Pi_{0}  \right)}{a^2 k^2}+\sigma\right)}-3 \sigma H}
\end{equation} 
In the limit $t \to 0$, we have \begin{equation}
    w = \frac{-\frac{\sqrt{\sigma \left(\frac{6 \left(a^2 H_{0}^2+ \Pi_{0}  \right)}{a^2 k^2}+ \sigma \right)} \left(a^2 \left(\sigma k^2+2 H_{0} (3 H_{0}+\lambda)\right)+6 \Pi_{0} \right)}{a^2 \left(\sigma k^2+6 H_{0}^2\right)+6 \Pi_{0}}+\sigma+3 \zeta_{0} H_{0}}{\sqrt{\sigma \left(\frac{6 \left(a^2 H_{0}^2+\Pi_{0} \right)}{a^2 k^2}+\sigma\right)}-\sigma}
\end{equation}
While in the far future limit $t \to \infty $ we have \begin{equation}
    w = -1 + \zeta_{0} k \sqrt{\frac{3 }{2 \sigma}}
\end{equation}
In the far future limit, even for the non flat space we have the universe tending towards a quintessence evolution, again something which we didnt observe in the Chern Simons theory. 
\\
\\
For the pseudo rip with $H(t) = H_{0} - H_{1} \exp(-\lambda t) $, we have the conservation law as 
\begin{multline}
3 \left(H_{0}-H_{1} e^{-\lambda t}\right) \left((w+1) \left(\sqrt{\sigma \left(\frac{6 \left(\frac{\Pi_{0}}{a^2}+\left(H_{0}-H_{1} e^{-\lambda t}\right)^2\right)}{k^2}+\sigma\right)}-\sigma\right) \right. \\
\left. \frac{2 \sigma \lambda H_{1} e^{-\lambda t}}{k^2 \sqrt{\sigma \left(\frac{6 \left(\frac{\Pi_{0}}{a^2}+\left(H_{0}-H_{1} e^{-\lambda t}\right)^2\right)}{k^2}+\sigma\right)}}-  3 \zeta_{0} \left(H_{0}-H_{1} e^{-\lambda t}\right)\right) = 0
\end{multline}
In the case of $\Pi_{0} = 0 $, we have \begin{multline}
    w = -1 + \frac{3 \zeta_{0} \left(H_{0}-H_{1} e^{-\lambda t}\right)}{\sqrt{\sigma \left(\sigma+\frac{6 \left(H_{0}-H_{1} e^{-\lambda t}\right)^2}{k^2}\right)}-\sigma}-  \frac{2 \sigma \lambda H_{1} e^{-\lambda t}}{k^2 \left(\sqrt{\sigma \left(\sigma+\frac{6 \left(H_{0}-H_{1} e^{-\lambda t}\right)^2}{k^2}\right)}-\sigma\right) \sqrt{\sigma \left(\sigma+\frac{6 \left(H_{0}-H_{1} e^{-\lambda t}\right)^2}{k^2}\right)}}
\end{multline}In the limit of the far future, we have \begin{equation} \label{eq31}
w = \frac{3 \zeta_{0} H_{0}}{\sqrt{\sigma \left(\sigma+\frac{6 H_{0}^2}{k^2}\right)}-\sigma}-1
\end{equation}
In this case, we see that as opposed to what happens in the conventional cosmology, one sees here that a quintessence phase of evolution is to happen for sure in the late universe. Furthermore, in the limit of $t \to 0$, we have \begin{equation}
w = -1 \frac{3 \zeta_{0} (H_{0}-H_{1})}{\sqrt{\sigma \left(\sigma+\frac{6 (H_{0}-H_{1})^2}{k^2}\right)}-\sigma}-\frac{2 \sigma \lambda H_{1}}{k^2 \left(\sqrt{\sigma \left(\sigma+\frac{6 (H_{0}-H_{1})^2}{k^2}\right)}-\sigma\right) \sqrt{\sigma \left(\sigma+\frac{6 (H_{0}-H_{1})^2}{k^2}\right)}}
\end{equation}
For the non-flat universe, we similarly have \begin{multline} \label{eq33}
    w = -1 + \frac{3 \zeta_{0} \left(H_{0}-H_{1} e^{-\lambda t}\right)}{\sqrt{\sigma \left(\frac{6 \left(\frac{\Pi_{0}}{a^2}+\left(H_{0}-H_{1} e^{-\lambda t}\right)^2\right)}{k^2}+\sigma\right)}-\sigma}-  \\ \frac{2 \sigma \lambda H_{1} e^{-\lambda t}}{k^2 \sqrt{\sigma \left(\frac{6 \left(\frac{\Pi_{0}}{a^2}+\left(H_{0}-H_{1} e^{-\lambda t}\right)^2\right)}{k^2}+\sigma\right)} \left(\sqrt{\sigma \left(\frac{6 \left(\frac{\Pi_{0}}{a^2}+\left(H_{0}-H_{1} e^{-\lambda t}\right)^2\right)}{k^2}+\sigma\right)}-\sigma\right)}
\end{multline}In the $t \to 0$ limit, we have \begin{multline} \label{eq34} 
     w = -1 + \frac{3 \zeta_{0} (H_{0}-H_{1})}{\sqrt{\sigma \left(\frac{6 \left(\frac{\Pi_{0}}{a^2}+(H_{0}-H_{1})^2\right)}{k^2}+\sigma\right)}-\sigma}- \\ \frac{2 \sigma \lambda H_{1}}{k^2 \sqrt{\sigma \left(\frac{6 \left(\frac{\Pi_{0}}{a^2}+(H_{0}-H_{1})^2\right)}{k^2}+\sigma\right)} \left(\sqrt{\sigma \left(\frac{6 \left(\frac{\Pi_{0}}{a^2}+(H_{0}-H_{1})^2\right)}{k^2}+\sigma\right)}-\sigma\right)}
\end{multline}
While in the far future, \begin{equation} \label{eq35}
    w = \frac{3 \zeta_{0} H_{0}}{\sqrt{\sigma \left(\frac{6 \left(\frac{\Pi_{0}}{a^2}+H_{0}^2\right)}{k^2}+\sigma\right)}-\sigma}-1
\end{equation}
Again, we see that a quintessence phase will occur in the pseudo rip case even for the non-flat universe. Finally, we have the $\zeta_{0}$ parameter as \begin{equation}
    \zeta_{0} = \frac{(w+1) \left(\sqrt{\sigma \left(\frac{6 \left(\frac{\Pi_{0}}{a^2}+\left(H_{0}-H_{1} e^{-\lambda t}\right)^2\right)}{k^2}+\sigma\right)}-\sigma\right)+\frac{2 \sigma \lambda H_{1} e^{-\lambda t}}{k^2 \sqrt{\sigma \left(\frac{6 \left(\frac{\Pi_{0}}{a^2}+\left(H_{0}-H_{1} e^{-\lambda t}\right)^2\right)}{k^2}+\sigma\right)}}}{3 \left(H_{0}-H_{1} e^{-\lambda t}\right)}
\end{equation}
\\
\\
For the power law model $a(t)=a_{0}+a_{1} (t-t_{0})^{2n}$ we have \begin{multline}
    (w+1) (t-t_{0})^2 \left(a_{0}+a_{1} (t-t_{0})^{2n}\right)^2 \left(\sqrt{\sigma \left(\frac{6 \left(4 a_{1}^2 n^2 (t-t_{0})^{4n}+\Pi_{0} (t-t_{0})^2\right)}{k^2 (t-t_{0})^2 \left(a_{0}+a_{1} (t-t_{0})^{2n}\right)^2}+\sigma\right)}-\sigma\right)- \\ \frac{2 \sigma \left(2 a_{1} n (t-t_{0})^{2n} \left(-2 a_{0} n+a_{0}+a_{1} (t-t_{0})^{2n}\right)+\Pi_{0} (t-t_{0})^2\right)}{k^2 \sqrt{\sigma \left(\frac{6 \left(4 a_{1}^2 n^2 (t-t_{0})^{4n}+\Pi_{0} (t-t_{0})^2\right)}{k^2 (t-t_{0})^2 \left(a_{0}+a_{1} (t-t_{0})^{2n}\right)^2}+\sigma\right)}} \\ - 6 a_{1} \zeta_{0} n (t-t_{0})^{2n+1} \left(a_{0}+a_{1} (t-t_{0})^{2n}\right) = 0
\end{multline}The equation of state parameter in this case comes out to be for $\Pi_{0} \neq 0$ \begin{multline}
    w = -1 + \frac{6 a_{1} \zeta_{0} n (t-t_{0})^{2n-1}}{\left(a_{0}+a_{1} (t-t_{0})^{2n}\right) \left(\sqrt{\sigma \left(\frac{6 \left(4 a_{1}^2 n^2 (t-t_{0})^{4n}+\Pi_{0} (t-t_{0})^2\right)}{k^2 (t-t_{0})^2 \left(a_{0}+a_{1} (t-t_{0})^{2n}\right)^2}+\sigma\right)}-\sigma\right)}+ \\  2 \sigma \left(2 a_{1} n (t-t_{0})^{2n} \left(-2 a_{0} n+a_{0}+a_{1} (t-t_{0})^{2n}\right)+\Pi_{0} (t-t_{0})^2\right) \\ \frac{1}{k^2 (t-t_{0})^2 \left(a_{0}+a_{1} (t-t_{0})^{2n}\right)^2 \left(\sqrt{\sigma \left(\frac{6 \left(4 a_{1}^2 n^2 (t-t_{0})^{4n}+\Pi_{0} (t-t_{0})^2\right)}{k^2 (t-t_{0})^2 \left(a_{0}+a_{1} (t-t_{0})^{2n}\right)^2}+\sigma\right)}-\sigma\right)} \\ \frac{1}{\sqrt{\sigma \left(\frac{6 \left(4 a_{1}^2 n^2 (t-t_{0})^{4n}+\Pi_{0} (t-t_{0})^2\right)}{k^2 (t-t_{0})^2 \left(a_{0}+a_{1} (t-t_{0})^{2n}\right)^2}+\sigma\right)}} 
\end{multline}In the limit of $t \to t_{0}$ we have \begin{equation} \label{eq39}
w = -1 + \frac{2 a_{0}^2 \Pi_{0}}{a_{0}^2 k^2 \left(\sigma-\sqrt{\sigma \left(\frac{6 \Pi_{0}}{a_{0}^2 k^2}+\sigma\right)}\right)+6 \Pi_{0}}
\end{equation}
Again we see that instead of the phantom regime as was seen in the case of the conventional cosmology, we have a quintessence regime happening instead. Also, we notice that for the flat universe $\Pi_{0} \to 0$, one finds that at the time of the bounce $w \to -1$ while in the simple cosmology w diverges in this scenario.  Finally for the sake of completeness we also write the viscosity parameter here to be \begin{multline}
    \zeta_{0} = \frac{(t-t_{0})^{-2n-1} \left((w+1) (t-t_{0})^2 \left(a_{0}+a_{1} (t-t_{0})^{2n}\right)^2 \left(\sqrt{\sigma \left(\frac{6 \left(4 a_{1}^2 n^2 (t-t_{0})^{4n}+\Pi_{0} (t-t_{0})^2\right)}{k^2 (t-t_{0})^2 \left(a_{0}+a_{1} (t-t_{0})^{2n}\right)^2}+\sigma\right)}-\sigma\right)\right)}{6 a_{1} n \left(a_{0}+a_{1} (t-t_{0})^{2n}\right)} - \\ \frac{-\frac{2 \sigma \left(2 a_{1} n (t-t_{0})^{2n} \left(-2 a_{0} n+a_{0}+a_{1} (t-t_{0})^{2n}\right)+  \Pi_{0} (t-t_{0})^2\right)}{k^2 \sqrt{\sigma \left(\frac{6 \left(4 a_{1}^2 n^2 (t-t_{0})^{4n}+\Pi_{0} (t-t_{0})^2\right)}{k^2 (t-t_{0})^2 \left(a_{0}+a_{1} (t-t_{0})^{2n}\right)^2}+\sigma\right)}}}{6 a_{1} n \left(a_{0}+a_{1} (t-t_{0})^{2n}\right)}
\end{multline}
For the exponential model we have $a(t)=a_{0} \exp \left(a_{1} (t-t_{0})^{2 n}\right)$, and for this we have the continuity equation as \begin{multline}
    6 a_{2} k^2 n (t-t_{0})^{2 n-1} \sqrt{\sigma \left(\frac{6 \left(\frac{\Pi_{0} e^{-2 a_{2} (t-t_{0})^{2 n}}}{a_{0}^2}+4 a_{2}^2 n^2 (t-t_{0})^{4 n-2}\right)}{k^2}+\sigma\right)} \\ \left((w+1) \left(\sqrt{\sigma \left(\frac{6 \left(\frac{\Pi_{0} e^{-2 a_{2} (t-t_{0})^{2 n}}}{a_{0}^2}+4 a_{2}^2 n^2 (t-t_{0})^{4 n-2}\right)}{k^2}+\sigma\right)}-\sigma\right)-6 a_{2} \zeta_{0} n (t-t_{0})^{2 n-1}\right)+ \\ 3 \sigma \left(8 a_{2}^2 n^2 (2 n-1) (t-t_{0})^{4 n-3}-\frac{4 a_{2} n \Pi_{0} (t-t_{0})^{2 n-1} e^{-2 a_{2} (t-t_{0})^{2 n}}}{a_{0}^2}\right) = 0
\end{multline}We have the equation of state parameter as \begin{multline} \label{nonflatw}
    w = -1 +\frac{6 a_{2} \zeta_{0} n (t-t_{0})^{2 n-1}}{\sqrt{\sigma \left(\frac{6 \left(\frac{\Pi_{0} e^{-2 a_{2} (t-t_{0})^{2 n}}}{a_{0}^2}+4 a_{2}^2 n^2 (t-t_{0})^{4 n-2}\right)}{k^2}+\sigma\right)}-\sigma}+ \\ 2 \sigma e^{-2 a_{2} (t-t_{0})^{2 n}} \left(\Pi_{0} (t-t_{0})^2-2 a_{0}^2 a_{2} n (2 n-1) (t-t_{0})^{2 n} e^{2 a_{2} (t-t_{0})^{2 n}}\right) \\ \Bigg[\frac{1}{a_{0}^2 k^2 (t-t_{0})^2 \left(\sqrt{\sigma \left(\frac{6 \left(\frac{\Pi_{0} e^{-2 a_{2} (t-t_{0})^{2 n}}}{a_{0}^2}+4 a_{2}^2 n^2 (t-t_{0})^{4 n-2}\right)}{k^2}+\sigma\right)}-\sigma\right)} \\ \frac{1}{\sqrt{\sigma \left(\frac{6 \left(\frac{\Pi_{0} e^{-2 a_{2} (t-t_{0})^{2 n}}}{a_{0}^2}+4 a_{2}^2 n^2 (t-t_{0})^{4 n-2}\right)}{k^2}+\sigma\right)}} \Bigg]  
\end{multline}
For the case of a flat universe \begin{multline}
    w = -1 + \frac{6 a_{2} \zeta_{0} n (t-t_{0})^{2 n-1}}{\sqrt{\sigma \left(\frac{24 a_{2}^2 n^2 (t-t_{0})^{4 n-2}}{k^2}+\sigma\right)}-\sigma}- \frac{4 a_{2} \sigma n (2 n-1) (t-t_{0})^{2 n-2}}{k^2 \left(\sqrt{\sigma \left(\frac{24 a_{2}^2 n^2 (t-t_{0})^{4 n-2}}{k^2}+\sigma\right)}-\sigma\right) \sqrt{\sigma \left(\frac{24 a_{2}^2 n^2 (t-t_{0})^{4 n-2}}{k^2}+\sigma\right)}} 
\end{multline}
For the flat universe scenario, when one goes to the time of the bounce $t \to t_{0}$, $w \to -1$ which is again in contrast to the conventional scenario but in the case of the non-flat universe \eqref{nonflatw}, we have $w \to \infty$ at the time of the bounce which is similar to the simple cosmology.

\section{Conclusions and final remarks}
We conclude our work with final comments and an overview on the implications of the analysis here. We can make the following observations : \begin{itemize}
    \item Let's consider firstly the case of the little rip and focus on \eqref{chernw} and \eqref{wrs2rip}. These are the equation of state parameters in the case of little rip for Chern Simons and RS-II cosmologies respectively and we notice that while in the case of chern simons a phantom evolution is supported for LR (as in the case of the GR cosmology as you discussed in your paper), in the case of RS-II a Quintessence scenario is supported. Going further, in the $t \to 0$  limit as well, Chern Simons supports phantom but in the $t \to \infty $ limit( far future), Chern Simons shows a pure dS evolution with w tending to -1 which is different than what happens in the conventional cosmology as you showed it goes to phantom even in the $t \to \infty$ limit. RS-II cosmology \eqref{wrs2farfuture} , meanwhile, supports only quintessence in both $t \to 0$ and $t \to \infty$ limit, no dS and no phantom, which is again a very fascinating fact. Also, this evolution scheme remains the same for both Chern Simons and RS-II in both flat and non flat universes, which is again surprising.
    \item In the case of the pseudo rip, a similar situation is observed. In both the cases of the flat and non flat universes, w tends to phantom regime for Chern Simons \eqref{eq13}, as it does for the usual cosmology while in both flat and non flat universes w tends to quintessence regime for RS-II braneworld (note how \eqref{eq35} reduces to the same as \eqref{eq31} in the flat universe limit ). An extra interesting bit is that in this case RS-II could allow for the Phantom regime too for the case in which t is not tending to infinity, as seen by \eqref{eq33} and \eqref{eq34}. So in both LR and PR, it's only in the pseudo rip scenario where for a non extreme limit the RS-II universe could go phantom but eventually in the far future it would again have to be in a quintessence evolution phase in the case of PR. 
    \item In the case of the bounce models, for the power law model RS-II again has quintessence evolution both in the limit of the bounce time as $t \to t_{0}$ \eqref{eq39} while in the conventional cosmology one observed phantom regime for the same case. The incredibly interesting thing that now occurs is that in this case even the Chern Simons cosmology has a quintessence evolution rather than a phantom evolution \eqref{eq18}, which is surprising seeing how Chern Simons was preferring phantom regime in the cases of rips. Both of these results hold true in both flat and non flat universes.
    \item In the case of the exponential model, for the flat universe Chern Simons goes phantom while for the case of the non flat universe it can go into Quintessence type evolution which is exactly the behaviour shown by the GR cosmology for the exponential model as you showed. For the flat universe, RS-II has a possibility of allowing for any of Quintessence or phantom types of evolution but for a non flat universe the RS-II cosmology goes completely quintessence which again distinguishes its behaviour from the usual cosmology for the same exponential model. 
\end{itemize}
All in all, the RS-II cosmology shows a stronger departure in the way it behaves under all of these scenarios than what happens in these scenarios in the usual GR cosmology. Chern Simons for a lot of cases shows similar behaviour to the conventional cosmology but also shows differences at various places. What is interesting is that both of these cosmologies are primarily higher dimensional theories ( in fact both being 5 dimensional in their usual forms ) and both have one quantity which distinguishes their features from the GR cosmology ( brane tension for the RS-II case and the $\alpha$ parameter for Chern Simons) but while being similar models in their makings, they respond to the cases of Little Rip, Pseudo Rip and Bounce cosmologies very differently. 

\section*{Acknowledgements}
The authors would like to thank Sergei Odintsov for various helpful discussions on singularities. This work was supported by Russian Foundation for Basic Research; Project No. 20-52-05009

\bibliography{citations}
\bibliographystyle{unsrt}

\end{document}